# Quantum Efficiency and Oscillator Strength of InGaAs Quantum Dots for Single-Photon Sources emitting in the Telecommunication O-Band

*Jan Große[1], Pawel Mwrowinski[1,2], Nicole Srocka[1] and Stephan Reitzenstein[1]*

[1]*Institut für Festkörperphysik, Technische Universität Berlin, Hardenbergstraße 36, 10623 Berlin, Germany*

[2]*Permanent address: Laboratory for Optical Spectroscopy of Nanostructures, Department of Experimental Physics, Wrocław University of Technology, Wybrzeże Wyspiańskiego 27, Wrocław, Poland*

**Abstract**

We demonstrate experimental results based on time-resolved photoluminescence spectroscopy to determine the oscillator strength (OS) and the internal quantum efficiency (IQE) of InGaAs quantum dots (QDs). Using a strain-reducing layer (SRL) these QDs can be employed for the manufacturing of single-photon sources (SPS) emitting in the telecom O-Band. The OS and IQE are evaluated by determining the radiative and non-radiative decay rate under variation of the optical density of states at the position of the QD as proposed and applied in J. Johansen et al. Phys. Rev. B 77, 073303 (2008) for InGaAs QDs emitting at wavelengths below 1 µm. For this purpose, we perform measurements on a QD sample for different thicknesses of the capping layer realized by a controlled wet-chemical etching process. From numeric modelling the radiative and nonradiative decay rates dependence on the capping layer thickness, we determine an OS of 24.6 ± 3.2 and a high IQE of about (85 ± 10) % for the long-wavelength InGaAs QDs.

**Introduction**

Semiconductor quantum dots (QDs) are well-established solid-state nanocrystalline structures that can be used as close-to-ideal two-level quantum emitters applicable for instance in quantum communication [1, 2, 3] or quantum computing [4, 5]. In recent years, performance of single-photon emission based on InGaAs QDs could be pushed towards the physical limits for instance in terms of photon indistinguishability up to 99.7% [6] for QD emitting in the 930-950 nm range. Moreover, by material engineering the range of applications could be extended to the telecom O-band at 1.3 µm [7, 8, 9, 10] which enables long-distance fiber-based quantum communication [11]. A very promising way to achieve the required redshift of QD emission is the introduction of a strain-reducing layer of lower Indium content on top of the QD layer [12] maintaining the high quality of emission in terms of multi-photon suppression [9]. An additional advantage of this technology route is the usage of a mature and well-established GaAs-based technology platform which facilitates the fabrication of advanced quantum light sources with enhanced photon extraction efficiency [13, 14, 15] and the possibility of spectral fine-tuning by external strain fields [16]. The QDs used in this work have been carefully optimized for the manufacturing of single-photon sources emitting in a wavelength range around 1.3 µm at temperatures up to 40 K as detailed in previously published works [15, 16, 17].

An important intrinsic property of any QD emitter is the internal quantum efficiency (IQE) which describes the ratio of quantum conversion from exciton formation into single-photon emission. The IQE plays a major role in the total quantum efficiency of a fully-processed device and directly impacts the photon extraction efficiency being a key parameter of quantum light sources. Thus, knowing and increasing the IQE is crucial for device optimization. Another important QD parameter is the oscillator strength (OS) of the optical transition. The OS is directly related to the radiative decay time of the emitter and determines the light-matter coupling strength of QD devices using effects of cavity quantum electrodynamics [18]. The exciton decay dynamics and IQE of In(Ga)As QDs emitting in the 920 to 1030 nm range have been experimentally examined by several groups [19, 20, 21, 22] with a method introduced by Johansen et al. [19] which allows for a direct measurement of the QD properties without being influenced by QD density or relying on assumptions based on theoretical considerations of the exciton radius and the dimensions of the QDs [23, 24]. This method was also used to investigate an InGaAs quantum dot - quantum well system [25]. So far however, to our best knowledge, no such measurements have been performed on InGaAs QDs emitting in the telecom O-band, so that reliable information on their IQE has not been available.

In order to determine the IQE and OS of telecom wavelength InGaAs QDs their total decay rate $\Gamma(\omega, z)$ at a frequency $\omega$ and at $z$ position below the sample surface is determined experimentally. The associated decay rate is given by the following relation:

$$\Gamma(\omega, z) = \Gamma_{nrad}(\omega) + \Gamma_{rad}^{hom}(\omega) \frac{\rho(\omega,z)}{\rho_{hom}(\omega)} \quad (1)$$

where $\Gamma_{nrad}(\omega)$ represents the nonradiative decay rate and $\Gamma_{rad}^{hom}(\omega)$ is the radiative decay in a homogeneous medium, $\rho(\omega, z)$ is the projected local density of states (LDOS) divided by the LDOS in a homogenous medium $\rho_{hom}(\omega)$ (the GaAs surrounding the QD). The total decay rate $\Gamma(\omega, z)$ in dependance from $z$ can be measured experimentally by means of time-resolved micro-photoluminescence (µPL) under variation of the capping layer thickness of the QD sample. Subsequently, the experimental components $\Gamma_{rad}(\omega)$ and $\Gamma_{nrad}^{hom}(\omega)$ are obtained by fitting parameters of a curve that has been derived from simulated values of the LDOS in dependance from $z$ to the experimental data. The IQE can then be calculated as:

$$IQE = \frac{\Gamma_{rad}^{hom}}{\Gamma_{nrad} + \Gamma_{rad}^{hom}}, \quad (2)$$

and the OS can easily be derived from

$$f_{osc}(\omega) = \frac{6 m_e \epsilon_0 \pi c_0^3}{q^2 n \omega^2} \Gamma_{rad}^{hom}(\omega) \quad (3)$$

where the electron mass $m_e$, the vacuum permittivity $\epsilon_0$, the vacuum speed of light $c_0$, the elementary charge $q$ and the refractive index $n$ of the surrounding GaAs is given [9].

**Sample Preparation**

The purpose of this work is to obtain a quantitative understanding of 1.3 µm InGaAs QDs, which were optimized for the development of quantum light sources emitting in the telecom O-band, in terms of their IQE and the OS. Using metal-organic chemical vapor deposition (MOCVD) the corresponding sample was prepared in the following way. First, a 300 nm GaAs buffer is grown on n-doped GaAs (100) substrate and followed by 1 µm thick $Al_{0.90}Ga_{0.10}As$ layer. Then a 2 µm thick GaAs layer and another 100 nm thick $Al_{0.90}Ga_{0.10}As$ layer is deposited. These three layers are originally designed as sacrificial layers to be removed during a post-growth flip-chip processing, as described in more detailed way in Refs. [15, 16]. The final 879 nm thick layer of GaAs formerly designed to function as active device membrane includes a single InGaAs QD layer together with the SRL.

The QD layer with a density of $5 \times 10^9$ cm$^{-2}$ is located 637 nm above the second etch stop layer and is formed by 1.5 monolayers of $In_{0.7}Ga_{0.3}As$ followed by GaAs flush corresponding to a nominal thickness of half a monolayer. The subsequent 5.5 nm thick InGaAs SRL has a gradual decrease of the In-content from 30% to 10% over the first 3.5 nm. This layer is followed by a GaAs capping layer with a thickness of 236 nm. An overview of the whole sample structure is given in Fig. 1(a). Further information on the QD growth and optimization can be found in the supplementary information.

In order to vary the thickness of the caping layer as required for extracting the radiative and non-radiative emission rate of the QDs the grown sample was prepared as described in the following. First, the surface of the sample was cleaned thoroughly and the photo resist AZ 701 MIR was applied using spin coating. UV lithography was used to pattern the surface with several slim rectangular structures to form a height reference to determine the capping layer thickness after the upcoming etching process. The sample was cut into 25 pieces to undergo wet-chemical etching after thorough cleaning with isopropanol. The etching solution [26] was prepared with the following ingredients: Hydrobromic acid (HBr), hydrogen peroxide ($H_2O_2$) and water ($H_2O$) in a volumetric ratio of 2:1:60. The solution was continuously agitated with a magnetic stirrer and left to rest for 600 seconds after adding the $H_2O_2$. The etching rate had been previously determined to be 0.5 nm/s by etching and measuring a set of test sample pieces of GaAs. The 25 pieces were put together into the solution and subsequently removed piece after piece after a set time interval to etch away the desired amount of GaAs capping and put into clean water to stop the etching process.

The remaining thickness of the GaAs separating the QD layer from the sample surface was determined with comparative measurement in a stylus profiler and an atomic force microscope (AFM) setup. Both measurements agree within a margin of less than 10 nanometers. Because of local deviations in the etching rate the exact position of the photoluminescence lifetime measurement is later marked on an optical microscopic image and the local top GaAs layer thickness remeasured via AFM. These values, as given in Fig. 1(b), are used in later analysis of the µPL data.

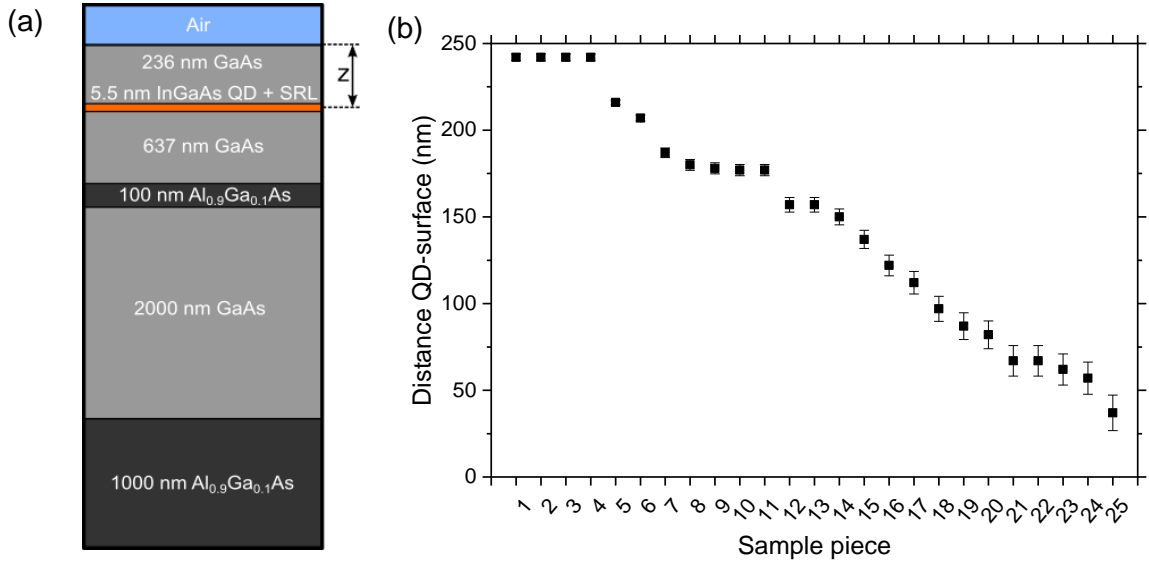

*Figure 1 (a) Epitaxial layout of the unprocessed sample. (b) Overview of the sample pieces (1-25) regarding the remaining GaAs top layer after etching as determined via AFM on the site of the optical measurements.*

**Optical measurements**

The optical characterization of the QDs was performed by means of micro-photoluminescence (µPL) spectroscopy at 10 K by mounting all samples with different top layer thickness in a He-flow cryostat. Optical excitation is provided by a continuous-wave diode laser (785 nm) and a tunable laser providing ps-pulses at a repetition rate of 80 MHz. Both lasers can be focused on the sample by a microscope objective with a numerical aperture (NA) of 0.4. The photoluminescence signal is collected with the same objective, spectrally resolved in a grating spectrometer (spectral resolution ~ 0.05 nm) and detected with a liquid nitrogen cooled InGaAs-array detector. Fig 2(a) displays a typical spectrum showing QD emission in the O-band on an unetched sample piece continuous wave excitation at 785 nm wavelength. The QD surface density of the wafer part used in the experiment is in the range of ~$10^9$ cm$^{-2}$. This is sufficient to identify single emission lines for a planar sample but still it means that with our spatial resolution of ~2 µm in diameter the µPL emission is collected from about 30-40 QDs which provides suitable averaging in the subsequent evaluation of the IQE and the OS of the QD ensemble.

For the subsequent time-resolved measurements the monochromator was set to a wavelength of 1260 nm at the center of the QD ensemble emission. This allows for measurements with suitable high signal-to-noise ratio which we will consider representative for the O-band range, even though it is located at the short-wavelength edge of this band. The photons are detected by a superconducting nanowire single photon detector (SNSPD) with a temporal resolution of ~50 ps and a detection efficiency of about 80% at ~1.3 µm. Fig 2(b) shows a typical decay curve of the excitonic QD states after each excitation pulse. Emission dynamics of QDs are usually modelled with a biexponential decay with a slow and a fast component, however only the fast decay is caused by bright excitons which are of interest regarding the IQE. After background correction the decay rate $\Gamma(z)$ was therefore obtained using a monoexponential fit of the fast decay. The LDOS $\rho_{hom}$ in dependance of z was calculated using the commercial finite-element simulation software JCMSuite.

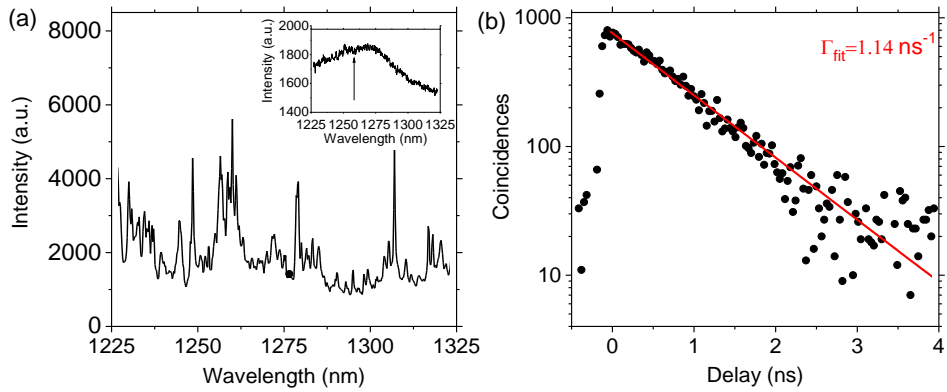

*Figure 2 (a) typical µPL spectrum in the telecom O-band range taken on the unprocessed sample at a temperature of 10 K and under continuous wave excitation at 785 nm with a power of 1.81 µW showing emission from multiple QDs. The inset shows a spectrum under pulsed excitation at 860 nm with a power of 300 µW used for TRPL measurements with the arrow marking the selected wavelength. (b) Exemplary time-resolved µPL measurement under pulsed excitation (80 MHz) and an exponential fit yielding a decay rate of $\Gamma$ =1.14 ns$^{-1}$.*

**Results and Discussion**

This section describes the extraction of the IQE and the OS based on the experimental results obtained by TRPL measurements for different top layer thicknesses. The corresponding data is depicted in Fig. 3. The experimentally obtained decay rates indicate an oscillating dependence which is related to the $\rho(z)$ factor in Eq. (1). In Fig. 3(a) the colored lines display the theoretical dependence derived from the simulation of the LDOS with a fixed $\Gamma_{rad}^{hom}$ of 1.15 ns$^{-1}$ and IQEs from 75% ($\Gamma_{nrad}$=0.38 ns$^{-1}$) to 95% ($\Gamma_{nrad}$=0.06 ns$^{-1}$). A comparison of the radiative and non-radiative decay rate components with results from Johannsen et al. and Albert et al. for standard InGaAs QDs and site-controlled InGaAs QDs at 930 nm, respectively, shows, that these numbers are similar to the range measured for In(Ga)As QDs without SRL with 1.0 ns$^{-1}$ for $\Gamma_{rad}^{hom}$ and 0.08 ns$^{-1}$ to 0.79 ns$^{-1}$ for $\Gamma_{nrad}$, which indicates that the present QDs have similar IQE. For the sake of clarity, in Fig. 3(b) measurements from pieces with similar etching depth are averaged which leads to a better agreement with the simulated curve and yields an IQE of (85 ± 10)% ($\Gamma_{nrad} = (0.20 \pm 0.14)$). This is similar if not quite as high as the reported value of 90% for standard self-assembled InAs QDs without SRL [19] and significantly higher than (0.47 ± 0.14) reported for nanohole-positioned InAs QDs [20]. We therefore conclude, that the SRL does not lead to a major increase in nonradiative decay rate from crystal defects or charge transfers to neighboring indium agglomerations which might form through segregation processes during the SRL overgrowth.

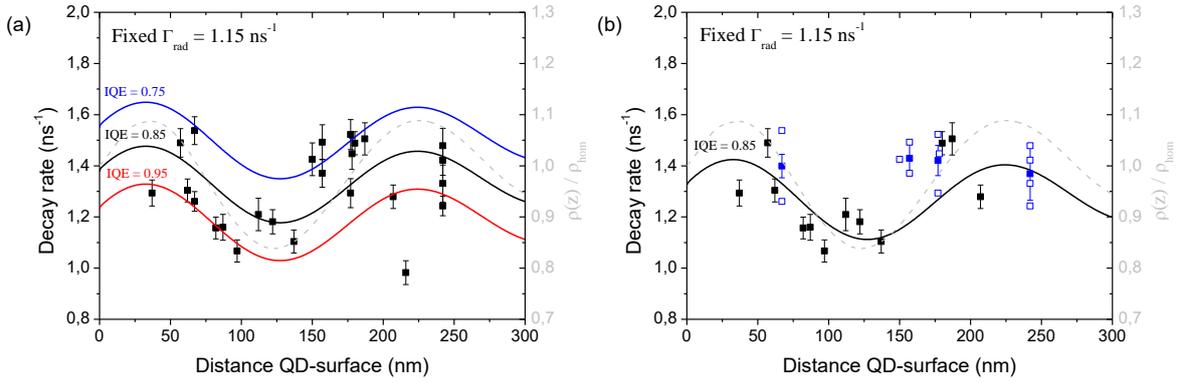

*Figure 3 (a) Measured (squares) and simulated (solid lines) emission decay rates of the processed QD samples with different top layer thickness. The simulated curves were calculated with a fixed radiative emission rate of $\Gamma_{rad}$ = 1.15 ns$^{-1}$) and with different assumed quantum efficiencies of IQE = 75 % (blue), IQE = 85 % (black) and IQE = 95 % (red). (b) Measured emission decay rates with average values (solid blue squares) calculated from measurements taken at similar etching depth (hollow blue squares). The hollow red square is omitted as a statistical outlier. In both plots the simulated lateral density of states $\rho(z)/\rho_{hom}$ is included (grey broken line) as a guide to the eye for the readers convenience.*

The upper limit of the OS for a strongly confined exciton in QDs which are smaller in size than the exciton Bohr radius is $f_{osc}^{max} = \frac{E_p}{\hbar\omega}$ [27, 22]. This yields a value of $f_{osc}^{max} = 25.5$ with a Kane energy of 25.1 eV [28] for our emitters derived from assuming 60% mean indium content in the QD which is lower than the nominal value but a realistic assumption due to known intermixing effects during QD growth and capping. The experimentally determined OS of our QDs, calculated according to Eq. (3) from $\Gamma_{rad}^{hom}$ using a low-temperature value for the refractive index of GaAs n = 3.34 [29] results in $f_{osc} = 24.6 \pm 3.2$. This is a high value for InGaAs QDs compared to previously reported data of 8 to 15 for structures emitting in the 930 – 950 nm range [19, 21, 20]. The increase in OS is not caused by a significant decrease in radiative lifetime but from an enlargement of the QD induced by incorporating additional indium during the SRL growth [30] leading to a lower ground state transition energy and a better overlap of the electron and hole envelope wavefunctions which is the main scaling factor of the OS in the strong confinement regime. The close agreement of the experimental value with the theoretical limit leads us to conclude that the envelope wavefunctions overlap in our QDs is close to unity. Interestingly, the high OS makes the long-wavelength QD interesting candidates for the study of cQED effects.

**Summary and Conclusion**

We performed an experimental evaluation of the OS and IQE of InGaAs QDs capped with an SRL that can be used to manufacture high quality telecom wavelength SPSs. For this purpose, the exitonic decay rates at the center of the QD ensemble emission band with a wavelength of 1260 nm were measured via TRµPL for different thicknesses from 242 to 37 nm of the GaAs top layer remaining after etching. By fitting the experimental results with the radiative decay rates and simulated LDOS data we found a radiative decay rate $\Gamma_{rad}^{hom}$ of $(1.15 \pm 0.15)$ ns$^{-1}$, a non-radiative decay $\Gamma_{nrad}$ of $(0.2 \pm 0.14)$ ns$^{-1}$. From this we calculated an IQE of $(85 \pm 10)\%$ and an OS of $24.6 \pm 3.2$, which confirms that the realized InGaAs QDs provide strong carrier confinement and keep their attractive properties for advanced quantum technologies when capped with an SRL.

Continuing the research should lead to further increase of the emission wavelength range. As our QDs feature emission range from 1200 to 1350 nm, a further investigation of the influence of QD size would be of strong interest. Also, measuring the temperature dependence of the OS up to 40 K would be insightful for practical applications such as stand-alone SPS cooled by Stirling cryocooler with based temperatures in the range of 35 – 40 K.

**Acknowledgements**

This work was funded by the FI-SEQUR project jointly financed by the European Regional Development Fund (EFRE) of the European Union in the framework of the program to promote research, innovation, and technologies (Pro FIT) in Germany within the 2nd Poland-Berlin Photonics Programme, Grant No. 2/POLBER-2/2016. The Authors also acknowledge support from the German Research Foundation through CRC 787 "Semiconductor Nanophotonics: Materials, Models, Devices" and Re2974/24-1, and the Volkswagen Foundation via project "NeuroQNet". P.M. gratefully acknowledges the financial support from the Polish Ministry of Science and Higher Education within the "Mobilnosc Plus–Vedycja" program and from the Polish National Agency for Academic Exchange (NAWA) via project PPI/APM/2018/1/00031/U/001. The authors thank T. Heindel for technical support financed via the German Federal Ministry of Education and Research (BMBF) via the project "QuSecure" (Grant No. 13N14876) within the funding program Photonic Research Germany.

**Data Availability**

The data that support the findings of this study are available from the corresponding author upon reasonable request.

**Supporting Information**

**Optimization of growth parameters for O-band-emitting QDs and SRL**

The MOCVD growth was performed in an AIXTRON 200/4 reactor under an ambient pressure of 100 mbar with standard precursor materials (TMGa, TMIn, AsH3 and TBAs). An overview on the primary growth parameters for the QD/SRL system can be found in Tab. I. The main goal of the optimization process was to achieve QD emission in the telecom O-band suitable for SPS processing by cathodoluminescence (CL) lithography. Therefore, the QD sheet density must be sufficiently small for individual lines to be identifiable, and the emission must be localized to an area corresponding to a single QD diameter. A major hurdle to overcome was the tendency of the QDs to cluster and couple leading to non-localized or broad ensemble peak emission. This was helped significantly by the introduction of a GaAs flush between the deposition of the QD layer and the SRL. Also, very low ratios of group V and group III materials (V/III ratio) in the reactor during the QD and SRL growth was found to be beneficial to the emission profile. Fig. 4 displays some examples of the change in QD emission in dependance from different parameter combinations.

*Table 1 showing the growth parameters and the respective range of variation investigated during the QD optimization process as well as the settled values used for the growth of the samples for device processing and the IQE and OS measurements.*

| Growth step | Parameter | Range | Settled value |
|---|---|---|---|
| QD layer | Indium content (%) | - | 66 |
| | Thickness (nm) | 1.5 – 2.3 | 1.5 |
| | V/III ratio | 0.5 – 2.8 | 0.5 |
| Growth interruption | Time (s) | 7.5 – 40 | 40 |
| GaAs Flush | Time (s) | 0.0 – 3.4 | 3.4 |
| | V/III ratio | 1.5 – 6.0 | 1.5 |
| SRL | In-content (%) | (33 – 48) to 10 gradient | 40 to 10 gradient |
| | Thickness (nm) | 2.7 – 5.5 | 5.5 |
| | V/III ratio | 1.5 – 6.0 | 1.5 |

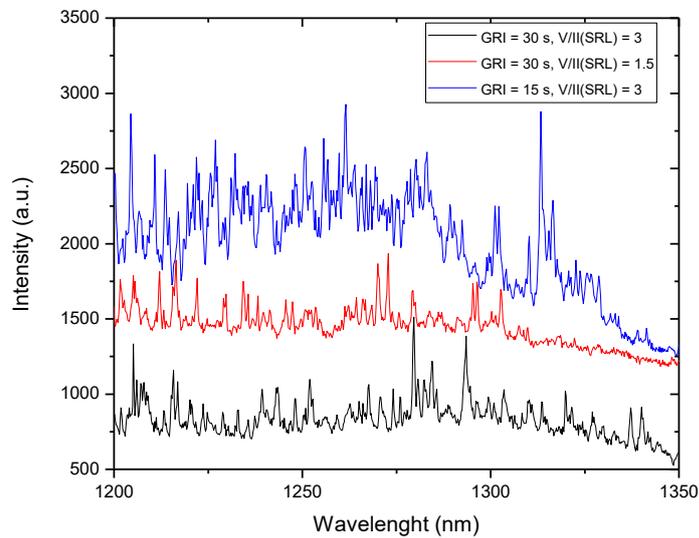

*Figure 4: Influence of changes in growth interruption (GRI) after QD layer deposition and V/III ratio during GaAs flush and SRL deposition on the emission spectrum of the samples. All other growth parameters are at their settled values as displayed in Tab. I.*